\documentclass[twoside,leqno,twocolumn]{article}  
\usepackage{ltexpprt} 
\usepackage[pdftex]{graphicx}
\usepackage{subfigure}
\usepackage{url}

\usepackage{algorithm}
\usepackage{algpseudocode}

\newtheorem{mydef}{Definition}

\usepackage{amsfonts,amsmath}

\renewcommand{\epsilon}{\varepsilon}

\newcommand{\R}{\mathbb{R}}

\DeclareMathOperator*{\argmax}{argmax}

\def\beglab{\begin{equation} \label}
\def\endlab{\end{equation}}
\def\beglabc{\begin{equation*} }
\def\endlabc{\end{equation*}}

\usepackage[usenames,dvipsnames]{xcolor}

\newcommand{\eg}{\textit{e.g.}}

\begin{document}


\title{Cold-start Problems in Recommendation Systems via Contextual-bandit Algorithms}
\author{
Hai Thanh Nguyen\footnote{INRIA Lille Nord Europe, 40 avenue Halley, 59650 Villeneuve d'Ascq, France, firstname.lastname@inria.fr}
\and 
J\'er\'emie Mary\footnote{University of Lille / LIFL (CNRS) \& INRIA Lille Nord Europe, 59650 Villeneuve d'Ascq, France, firstname.lastname@inria.fr}
\and
Philippe Preux\footnotemark[2]
}
\date{}

\maketitle
 

\begin{abstract}
In this paper, we study a cold-start problem in recommendation systems where we have completely new users entered the systems. There is not any interaction or feedback of the new users with the systems previoustly, thus no ratings are available.  Trivial approaches are to select ramdom items or the most popular ones to recommend to the new users. However, these methods perform poorly in many case. In this research, we provide a new look of this cold-start problem in recommendation systems. In fact, we cast this cold-start problem as a contextual-bandit problem. No additional information on new users and new items is needed. We consider all the past ratings of previous users as contextual information to be integrated into the recommendation framework. To solve this type of the cold-start problems, we propose a new efficient method which is based on the LinUCB algorithm for contextual-bandit problems. The experiments were conducted on three different publicly-available data sets, namely Movielens, Netflix and Yahoo!Music. The new proposed methods were also compared with other state-of-the-art techniques. Experiments showed that our new method significantly improves upon all these methods.
\end{abstract}

\section{Introduction}

The goal of a recommender system is to select some items (\eg{} movies,
music, books, news, images, web pages and so on) that are likely to be
of interest for a user on a webpage. This is done by matching some
user-based characteristics with item-based characteristics. These
characteristics can be information over the items to recommend (the
content-based approach) or to find users with similar tastes (the
collaborative filtering approach). 
   We consider the quality
of a recommendation to be the interest of the user on the item being
recommended.

In the content-based approach, a description of the items is readily
available. We have to build a model of the user's tastes using the
feedback we receive --- this feedback can be implicit (\eg{} clicks,
browsing, \ldots) or explicit (\eg{} rates). When a new user ---
without any side information --- is introduced in the system, we need
to collect some data in order to build a good enough model before
being able to produce any valuable recommendation. This is a first
kind of cold-start problem that we qualify as the \emph{new user
  problem}. In this setting we want to balance the exploration of the tastes of the new user and the usage of this modeling to perform good recommendations.  

In the collaborative filtering approaches, recommendations are made
trying to identify users with similar preferences. Of course, it
suffers from the new user problem, but it also suffers from a
\emph{new item problem}. Indeed, if a new item is introduced, the
system will not recommend it until some users provide some positive
feedback on it. Certainly, the probability of receiving some feedback
is related to the number of recommendations of that item, so it sounds
reasonable to provide a \emph{boost} to the new items. Of course, this
boost has to be designed carefully. If the boost is too strong, it will display too much the new items or even it will not be adapted to the current user (but we will have a lot of information on theses new items). If the boost is too low, the new items will never be displayed or even if they would be better than the old ones. So we  have to set a balance between the exploration of theses new items and the quality of our recommendations. 

Cold-start problems naturally arise in statistical data modeling when
not enough data is available. A common way to solve the new item
problem is to assign a default rate to new items based on the
ratings assigned by the community to other similar items \cite{Agarwal:2009:SME:1526709.1526713}. Item
similarity is computed according to the items content-based
characteristics. Certainly, this only holds when content-based
characteristics are available.




In this paper, we make no use of any content-based information on users and items. We propose a new approach based on the bandit problem to tackle both the new user and the new item problems. In fact, we cast the cold-start problems into contextual-bandit problems. We consider all the past ratings of previous users as context information in the recommendation systems. We adapt the standard LinUCB algorithm to our cases and propose a new efficient method, namely A-LinUCB algorithm.

To verify the proposed algorithm, we conduct an experiment on three data sets from Movielen, Netflix and Yahoo!Music. We also compare the new algorithm with different approaches, which are the random policy, $\epsilon$-greedy, UCB, EXP3 and Thompson sampling. Experimental results showed that our new method significantly improves upon all these methods in both new user and new item recommendation systems.

The paper is organized as follows. Section 2 provides definitions of the new user and new item problems. In the Section 3, we describes the bandit algorithms and their perspectives for cold-start problems. Section 4 introduces the contextual-bandit approach and an adapted version of the LinUCB algorithm for the cold-start problems. We present the experiments in Section 5. The last section summarizes our findings and discuss about future works. 





\section{Problem Definition}
\label{def}
Assume that at a given moment, we have a set of $n$ possible items to recommend. Let $X \in \R^{k\times n}$ be a matrix of description of theses items (one item per column). 

For all $i \in \{1,\ldots,m\}$, let $\theta_i \in \R^{k}$ be a row vector describing the $i^{th}$ user, and let $b(\theta_i) \in \R^n$ be a vector of tastes of this user for the $n$ items, a taste being expressed as a number. The $j^{th}$ value of $b(\theta_i)$ is the affinity of the user $i$ for the $j^{th}$ item. The affinity is built using implicit of explicit feedback over the tastes of the users. Some of the values are unknown. 

Let $U$ be the matrix of description of all users. The $i^{th}$ row of $U$ is $\theta_i$. 
Let $B \in \R^{m \times n}$ be the affinity matrix. The $i^{th}$ row of $B$ is the vector $b(\theta_i)$.

A common goal is to predict the missing values of $B$ using the available descriptions of users and/or items. 
Classically the error of an algorithm is seen as the reconstructing error of $B$ --- the users tastes --- from the available data. For that reconstruction, a classical measure of performance is the root mean square error (RMSE) but some authors have proposed different criterions such as rank preservation \cite{Shi:2012:CLM:2365952.2365981}.

To reconstruct the missing values, a very common assumption is to consider that there exists a latent description space common to both items and users, so that a linear relation holds between the tastes and the vectors of descriptions. Formally for each user we want to write $b(\theta_i) = \theta_i \cdot X$. When performed for all users at once, this method is known as matrix factorization since we are trying to write $B = U \cdot X $.

But this approach fails short to deal with the occurrence of new users or new items. Indeed, for rows or columns of $A$ with no or almost no information in the data set, there is no way to know if our reconstruction is correct or not.
In a real application, data are received in an online way: after each recommendation to an user, we have the possibility to collect a feedback. This means that it is possible to recommend items in order to collect information on the user (so as to categorize it) in order to better recommend latter. In particular, we are going to show that it is not optimal to always recommend the ``best'' item according to the current estimates of the tastes of the user, a strategy said to be \emph{greedy}. 
 
As in this online setting the RMSE evaluation is not enough, we have to rely on a different evaluation process.
Assuming a brand new user is visiting us, a reasonable recommender system tries to present the item with the highest affinity. 
One of the most common online evaluation is the regret, which is the deference of performance between making the best decision and the decision that has been made. Cumulative regret is simply the sum over time of the regret. The behavior of this quantity is studied in the bandits framework \cite{Auer:2002:FAM:599614.599677, DBLP:conf/nips/Abbasi-YadkoriPS11}.  
In fact, the cumulative regret is similar to the average score of our recommendations when done sequentially. The regret is more used than the average score because it highlights better the difference between algorithms --- this is because the average score of the items does not appear in the regret formulation so cumulative score grows in $O(T)$ while differences between algorithm will be in $O(\sqrt(T))$ where $T$ is the number of recommendations made ---Translated in our setting this means at each time step $t$:

\begin{enumerate}
  \item we receive the visit of a user $i_t$ who is either already
    known to the recommendation system, or not. For this user, at
    timestep $t$ some of her tastes $b(\theta_{i_t})$ are known by the
    recommender and some others are unknown. We note $b^*_t$ the highest
    $b(\theta_{i_t})$ currently not known by the recommender . Of course $b^{*}_{t}$ is not revealed, it is just used for the sake of the definition of the regret. 
  \item The recommender chooses an item to recommend $j_t =
    \pi(i_t,t)$ among the unknown one for $i_t$. The corresponding value of $b(\theta_{i_t})$ is
    revealed. The regret is increased by $(b^{*}_{t}-b_{\pi(i_t,t)}(\theta_{i{_t}}))$.
\end{enumerate}

The objective is to find a best strategy that provides the minimal cumulative regret:

$$ 
  CumulativeRegret = \sum_{t=1}^{T}(b^{*}_{t}-b_{\pi(i_t,t)}(\theta_{i{_t}}))
$$

Of course, the computation of the cumulative regret requires the knowledge of all the values of $B$ which can be problematic on some real datasets.

\subsection{New User/Item Problem}
\label{prob}





\begin{mydef}[The new user problem]
When a set of new users is introduced in our recommender system, we want to recommend the items to them and get the feedback on the recommended items minimizing the cumulative regret --- see section \ref{def} --- on theses users. 
\end{mydef}

Our global strategy for solving this problem is following:
\begin{enumerate}
	\item Select $k$ users. The tastes of theses users are seen as the descriptions of the items. This means that we use as $X$ matrix some of the rows of $B$. This special set of users is designed as the \emph{base users} and the $X$ is the \emph{base matrix}.
	\item As the description of the items contains some missing values, we are going to fill them using different strategies.
	\item Then select items using a contextual bandit algorithm as described in section \ref{linucb}. 
\end{enumerate}
We are going to find how to express the tastes of our new users as a linear combination of the base users. This is equivalent to find the $\theta_i$ parameters for this new users. When searching this combination we pay attention to the confidence intervals of our estimates and sometimes sample in order to reduce the variance of the estimates instead of selecting the optimal choice with respect to the current maximum of likelihood. 

\begin{mydef}[The new items problem]
When a set of new items is introduced in our recommender system, we want to use the visits of our users in order to improve our statistical confidence on theses items. We perform this allocation trying to minimize the cumulative regret on the users.  
\end{mydef}

Our global strategy for this case is almost the same with the new user problem's strategy. The only difference is that we will use the \emph{base users} to compute a confidence bound over the description of the items. 

In the next section, we will introduce the bandit algorithms and their perspectives to solve these cold-start problems.

\section{Bandit algorithms and perspectives for cold-start problems}
\label{MAB}


At each moment, we have to select an item to recommend to the present user. This selection is based on uncertain information. Therefore, we may either choose an item that is most likely to appeal to the user (based on uncertain grounds, though), or choose an item that is likely to bring new information. This means that we may either suggest an item that will bring immediate benefit, or select an item that will let the system improve in the longer run. This is the usual dilemma between exploitation (of already available knowledge) \textit{versus} exploration (of uncertainty), encountered in sequential decision making under uncertainty problems. This problem has been addressed for decades in the bandit framework that we briefly introduced now.

Let us consider a set of possible items. Each item is associated to an unknown probability of ratings. The game is repeated and the goal is to accumulate the minimal cumulative regret. In the present context, ratings may be binary (click/no-click), or belong to a set of values (a set of possible rate, ranging from 1 to 5). This problem has been studied for decades, many approaches have been proposed. Let us introduce a few of them that we use later in this paper:

\begin{itemize}
  \item \textbf{Random} consists in picking up one of the possible items, uniformly at random.
  \item \textbf{$\epsilon$-greedy(EGreedy)}~\cite{Auer:2002:FAM:599614.599677} consists in picking up the item that is currently considered the best with probability $\epsilon$ (exploit current knowledge), and pick it up uniformly at random with probability $1-\epsilon$ (explore to improve knowledge). Typically, $\epsilon$ is varying along time so that the items get greedier and greedier as knowledge is gathered.
  \item \textbf{UCB}~\cite{Auer:2002:FAM:599614.599677} consists in selecting the item that maximizes the following function: $\hat\mu_j + \sqrt{\frac{2\ln t}{t_j}}$ where $t$ is the current timestep, $\mu_j$ is the average rating obtained when selecting item $j$, $t_j$ is the number of times item $j$ as been selected so far. In this equation, $\hat\mu_j$ favors a greedy selection (exploitation) while the second term $\sqrt{\frac{2\ln t}{t_j}}$ favors exploration driven by uncertainty: it is a confidence interval on the true value of the expectation of rating for item $j$.
 


  \item \textbf{EXP3}~\cite{DBLP:journals/siamcomp/AuerCFS02} consists in selecting an item according to a distribution, which is a mixture of the uniform distribution and a distribution that assigns to each item a probability mass exponential in the estimated cumulative ratings for that item. 
  \item \textbf{Thompson sampling(TS)}~\cite{DBLP:journals/corr/abs-1209-3352} is a Bayesian approach to this problem. It consists in computing the probability of success of each recommendation, and then greedily selecting the item associated with maximum \textit{a posteriori} probability.
\end{itemize}


In the setting we consider in Section~\ref{def}, we have information about the base items/users and we may also have information about the ratings of the base matrix $X$. This is known as side information, or context, hence bandit with side information or contextual bandit. The bandit approaches considered so far do not take this side information into account. There are various methods for the contextual bandits (\eg{} OTS~\cite{May-et-al:2012} and LinUCB~\cite{LinUCB}); here, we concentrate on LinUCB~\cite{LinUCB} because of its efficiency. More details are given in the section \ref{linucb}. 
%
%
%
%
%
%



\section{Contextual-bandit for cold-start problems}
\label{linucb}
In the setting considered in this paper, we assume that there is no contextual information available about neither the items, nor the users. However, the ratings already recorded from users on items may be used as a context for the new users. We elaborate on this idea in the subsequent section.

We assume that there is a base rating matrix $X=(X_{j})_{j=1}^{n}$ of dimension $k\times n$, where $X_{j}$ is the $j^{th}$ column of $X$. It is not necessary that the $X$ to be a full rating matrix. Because in practice of recommendation systems, it is difficult to get all the ratings of the users for all available items. However, each user should have at least a rating. The missing values in $X$ can be filled by zero, or average values or the values approximated by using matrix decomposition techniques, such as Singular Value Decomposition (SVD). We will discuss about this more in the experiment part.

At a time step $t$, if a new user comes, her rating for a particular item $j$ is linear in its context vector $X_{j}$ with some unknown coefficient vector $\theta_{i_{t}}$:
$$
b_{j}(\theta_{i_{t}}) = \theta_{i_{t}}X_{j}
$$


Let $D_{t,j}$ be a context description matrix of dimension $t\times k$, where rows are the context vector $X_{j}$. When applying the ridge regression to the training data $(D_{t,j},b_{j})$, we get the estimation of the unknown variable $\theta_{i_{t}}$ as follows:
$$
\theta_{i_{t}} = (D^{T}_{t,j}D_{t,j}+I_{k})^{-1}D^{T}_{t,j}b_{j}
$$
where $I_{k}$ is the $k\times k$ identity matrix. As shown in~\cite{LinUCB}, with probability at least $1-\delta$ we have:
$$
|\theta_{i_{t}}X_{j}-E[b_{j}(\theta_{i_{t}})|X_{j}]|\leq \alpha\sqrt{X^{T}_{j}(A_{t,j})^{-1}X_{j}}
$$
for any $\delta > 0$, $\alpha=1+\sqrt{ln(1/\delta)/2}$; where $A_{t,j} = D^{T}_{t,j}D_{t,j}+I_{k}$ and $E[b_{j}(\theta_{i_{t}})|X_{j}]$ is the expected rating value of the new user on the item $j$, given the $X_{j}$. It can be easily seen that: $D^{T}_{t,j}D_{t,j}+I_{k} = tX_{j}X^{T}_{j}+I_{k}$. Therefore, 
$$
|\theta_{i_{t}}X_{j}-E[b_{j}(\theta_{i_{t}})|X_{j}]|\leq \alpha\sqrt{X^{T}_{j}(tX_{j}X^{T}_{j}+I_{k})^{-1}X_{j}}
$$

Obviously, we can immediately apply the standard LinUCB algorithm for this case. However, the efficiency of the algorithm will decreased over the time. Because when the size of $D_{t,j}$ increases, the inversion of the matrix $A_{t,j}$ is harder and slower. It is desirable to get more efficient algorithm than the standard LinUCB. Below, we propose a new adapted LinUCB (A-LinUCB) for the new user/item recommendation system problem and we will demonstrate its performance in the experiment part. Also note that unlike the standard LinUCB we work on context defined by rates and do not try to discover the \emph{best} item as we do not want to recommend an item with a known taste.

\begin{lemma}
  For all $t\geq 1$ and $X_{j} \in [0,1]^{k}$, the following inequation holds:
  $$
    (tX_{j}X^{T}_{j}+I_{k})^{-1} \leq (X_{j}X^{T}_{j}+I_{k})^{-1}
  $$ 
\end{lemma}

Following the lemma, we have a new inequality for A-LinUCB as follows:
$$
|\theta_{i_{t}}X_{j}-E[b_{j}(\theta_{i_{t}})|X_{j}]|\leq \alpha\sqrt{X^{T}_{j}(X_{j}X^{T}_{j}+I_{k})^{-1}X_{j}}
$$
The inequality leads to a new adapted LinUCB (A-LinUCB) as described in Algorithm~\ref{algo:LinUCB-for-new-user-item-pb}.

As the confidence interval $(X^{T}_{j}(X_{j}X^{T}_{j}+I_{k})^{-1}X_{j})$ for each item does not change over time, it may be a question about the impact of the exploration on the actual recommendation results. In other words, what if we do not explore different items by setting $\alpha$ to zero. It turns out that in practice we still need a little exploration because when $\alpha=0$, the performance of the algorithms will be dramatically decreased. The details of this discussion will be provided in the experiment part. 

\begin{algorithm}
	\label{AlinUCB}
  \caption{A-LinUCB for new user problem. (A-LinUCB for the new item problem is almost the same, except that the base matrix $X$ needs to be transformed before running the algorithm)} 
  \begin{algorithmic}[1]
    \Procedure{A-LinUCB}{$T,\alpha,X$}
    \State Store the matrices $A_{j} = I_{k}+X_{j}X^{T}_{j},j=1\ldots n.$
    \For {$t$ in $1:T$}
      \For {$j$ in $1:n$}
        \If {$j$ is new}
	  \State $b_{j} \gets 0_{n\times 1}$	
	\EndIf
	\State $\theta_{t}\gets{}A^{-1}_{j}b_{j}$
	\State $p_{t,j} \gets \theta_{t}X_{j}+\alpha\sqrt{X^{T}_{j}(A_{j})^{-1}X_{j}}$
      \EndFor
	  \State Draw a user at random $i_t$ 
      \State Choose item \\ ~ $j_{t}= \argmax_{j}\{p_{t,j} | \mbox{ taste of  $i_t$ for  j is unknown } \}$ 
      \State Observe the real rating $b_{j_{t}}$
	  \State $b_{j} \gets b_{j}+b_{j_{t}}X_{j_{t}}$	  	
    \EndFor
  \EndProcedure
  \end{algorithmic}
  \label{algo:LinUCB-for-new-user-item-pb}
\end{algorithm}
\section{Experiment}
In this section, we conduct comprehensive experiments to evaluate the performance of the proposed A-LinUCB on solving the cold-start problems in recommendation systems. In order to do that, we first give the detail on the data sets used in the experiments. We then describe the experimental settings, in which the implementation of the A-LinUCB and other methods to compare with are provided . Finally, we analyze and discuss about the experimental results.

\subsection{Data sets}
~\\
We used three different publicly-available data sets, namely Movielens\footnote{http://www.grouplens.org/}, Netflix\footnote{http://www.netflixprize.com/} and the Yahoo!Music\footnote{http://kddcup.yahoo.com/}. The data sets Netflix and Yahoo!Music contain users that have not any ratings for any items. We eliminated these users from the original data sets because with these users, it would not be possible to estimate the regret. In more detail, originally the Netflix data set contains 100,000 ratings (from 0 to 5, where 0 = not rated) of 68,357 users on 17,770 movies. After the elimination, we obtained a data set that has 13,545 ratings of 6423 users on 1250 items. With the original Yahoo!Music data set, we have 11,557,943 ratings of 98,111 artists by 1,948,882 anonymous users. We reduced the data to a smaller one, which contains 20,361,089 ratings of 50,080 artists by 483,273 users. Because of the limited memory, from the original eliminated Yahoo!Music dataset we randomly selected several subsets for our experiment and then the results will be averaged. In Table~\ref{table:datasets}, we summarize the size of data sets in our experiments. 

For the convenience in analysis of the results and in presenting the cumulative regrets on the graphs, we normalized the data sets before running the experiments. 

With each data set, we randomly splited it into two parts: The first part is the base matrix $X$ and the second part $Y$ is the remained data. Note that we will fill the missing values in the $X$ only and use the matrix $Y$ to measure the performance of the algorithms.
\begin{table}
  \caption{Number of users, items and ratings in datasets after eliminating the users without any votes from original datasets.}
  \centering
  \begin{tabular}{|l|c|c|c|}
    \hline
    Datasets & No. Users & No. Items & No. Ratings \\
    \hline
    Movielens & 6,040 & 3,952 & 1,000,209\\
    \hline
    Netflix & 6,423 & 1,250 & 13,545 \\
    \hline  
    Yahoo!Music & 483,273 & 50,080 & 20,361,089 \\
    \hline
  \end{tabular}
  \label{table:datasets}
\end{table}

\subsection{Experiment setting}
~\\
This subsection gives a detailed description of our experimental setup, including missing values handling, A-LinUCB algorithms for the new user and new item recommendation systems, competing algorithms and performance evaluation.

\subsubsection{Dealing with missing values}
~\\
It happens always in recommendation systems that some users do not give the rates on many items. Therefore, the data sets usually contain many missing values. In our experiments, the base matrix $X$ may have these values as well. It is necessary to fill the matrix $X$ before running the A-LinUCB algorithm. 

\begin{itemize}
\item The simplest approach is to use the zero value. This method is reasonable since in some cases, if an user does not rate for a particular item, then it is probably that she does not like the item. Of course, some items are unknown to the user. However, considering the efficiency of this approach, especially, when we will be able to use the computing advantage with large sparse matrix, we would take into account it and compare with other possible techniques.

\item The second approach, which is also very efficient in practice, is to utilize the average values of the ratings for items to fill the missing values. 

\item Recently, several matrix decomposition/completion techniques are applied and demonstrated to be promising approaches. In our experiment, we implement two popular algorithms, which are 1) the imputed SVD and 2) ALS-WR decomposition.

\textit{Ad.1} Firstly, the missing values in a column vector of $X$ are filled with the average value. The obtained matrix is then approximated by the standard SVD method~\cite{Billsus:1998:LCI:645527.657311} keeping only the $k$ biggest singular values. \\

\textit{Ad.2} ALS-WR~\cite{Zhou:2008:LPC:1424237.1424269} which is a Tikhonov regularization of the classical matrix decomposition with low rank assumption. The minimization is solved by a well initialized alternate least square and is very efficient on Netflix dataset. 

\end{itemize}

Three above described approaches are compared with each other in terms of the performance of the A-LinUCB. We will use the best result for later experiments.

\subsubsection{Settings of the A-LinUCB algorithms}
The implementation of the A-LinUCB algorithm is simple as shown in Algorithm~\ref{algo:LinUCB-for-new-user-item-pb}. The only parameter of this policy is the $\alpha$. We executed the A-LinUCB with different values of $\alpha$ and recorded the best choice. 

Basically, the A-LinUCB algorithms for the new user problem is different from the A-LinUCB for the new item problem. However, their implementations as shown in Algorithm~\ref{algo:LinUCB-for-new-user-item-pb} are almost the same, except that the selection of the base matrix $X$ of dimension $k\times n$ is distinct. The meanings of the $k$ and $n$ will be exchanged when we change from a problem to another problem. In other words, for the new item problem, the $k$ will become the number of base items and the $n$ will become the number of base users. 

In general, the values of $k$ and $n$ of the matrix $X$ can be arbitrary numbers. In our experiment, we observed that the case when $k$ equals to $n$ ($k=n$) provided the most consistent results and because of the limited size we only reported these results.

In our experiment, we selected the base matrix $X$ for each problem on each data set as follows:
\begin{itemize}
\item For the new user problem, with the Movielens and Netflix data sets we used the initial numbers of items for the dimension of the base matrices $X$. In more detail, the $X$ for the Movielens and for the Netflix will have the dimensions of $3952$ and $1250$, respectively. With the Yahoo!Music data set, because of the limited memory we randomly chose the base matrix $X$ of the size $1000\times 1000$ five times. The results were averaged.
 
\item For the new item problem, we picked up the matrices $X$ with following dimensions: $1000\times 1000$ (Movielens), $500\times 500$ (Netflix) and $1000\times 1000$ (Yahoo!Music).
\end{itemize} 

\subsubsection{Competing algorithms}
~\\
To the best our knowledge, no existing work applies contextual multi-armed bandit algorithms to solve the cold-start problems in recommendation systems when the side information on users and items is not available. Nevertheless, we compare the new proposed A-LinUCB with the following algorithms:

\begin{itemize}
\item The first and the most naive approach for the cold-start recommendation systems is to choose randomly an item to recommend to a new user. This algorithm is very efficient, especially, when we do not have any description on users or items, the algorithm seems to be reasonable. However, its performance in many cases fails as we will show in the experimental results. It is desirable to design better algorithms. 

\item The next class of algorithms that we want to compete with is the zero-contextual multi-armed bandit algorithms. The purpose of this comparison is to show the value of taking into account context information in the cold-start recommendation systems. The algorithms involved in the comparison are: 
\begin{itemize}
\item $\epsilon$-greedy(EG): As described in Section~\ref{MAB}, it estimates the rating of each user/item; then selects a random user/item with probability $\epsilon_{t}$, and choose an user/item of the highest average value of revealed ratings with probability $1-\epsilon_{t}$. The parameter $\epsilon_{t}$ is decreased over the time $t$. In fact, the $\epsilon_{t}$ is calculated as follows: $\epsilon_{t}=min(1,(cn)/(d^{2}(t-n-1)))$, where $c$ and $d$ are chosen constants.
\item UCB: As described in Section~\ref{MAB}, this policy estimates the average value of previously revealed ratings for each item as well as a confidence interval of the estimation. Afterward, the UCB always choose the item with the highest UCB. Specifically, following UCB~\cite{Auer:2002:FAM:599614.599677}, we calculated an item $j$'s confidence interval by $\sqrt{\frac{2\ln t}{t_j}}$, where $t_{j}$ is the number of times the item $j$ was selected prior to trial $t$.
\item EXP3: As described in Section~\ref{MAB} and accodring to the algorithm in~\cite{DBLP:journals/siamcomp/AuerCFS02}, we selected $\gamma=0.01$ before drawing the probability to select the best item to recommend to the new user.  
\end{itemize}

\item Thompson Sampling(TS) for contextual bandits: This policy attempts to utilize the previous ratings as the context information for the cold-start recommendation systems. By means of this comparison, we want to show that the utilization of the context information should always be in an appropriate way. The details on the implementation of the TS algorithm is as follows: At every step $t$, we generated a $n$-dimensional sample $\mu(t)$ from multi-variate Gaussian distribution, which depends on the true rating of the last time recommended item and its context information. We then solved the problem $argmax_{i}(\mu(t) X_{j})$ to get the next item for the recommendation.


\item Finally, since the confidence interval $X^{T}_{j}(X_{j}X^{T}_{j}+I_{k})^{-1}X_{j}$ described in Section~\ref{linucb} is fixed for each item, we doubt about the impact of the exploration in the cold-start recommendation systems. To answer the question, we compare the A-LinUCB with the A-LinUCB where the $\alpha$ is set to zero. For this case, at every time step $t$ an item with maximal value $\theta_{t}X_{j}$ is selected.

\end{itemize}

We ran each of these algorithms 10 times with different choices of parameters. The best results were recorded as shown in Table~\ref{table:newUser}, Table~\ref{table:newItem} and Table~\ref{table:A-LinUCB(0)vsA-LinUCB(0.001)}.

\subsubsection{Performance metric}
~\\
As discussed in Section~\ref{def}, it will not be able to use the offline evaluation methods, such as RMSE or MSE for the cold-start recommendation systems. In our experiments, we utilize an online measurement, which is the cumulative regret as defined as follows:
$$ 
  CumulativeRegret = \sum_{t=1}^{T}(b^{*}_{t}-b_{\pi(i_t,t)}(\theta_{i{_t}}))
$$
For this evaluation, the $b^{*}_{t}$ of the new user may be not available. Therefore, we will use the maximal value among known ratings of the user instead. We assume that the real rating value $b_{\pi(i_t,t)}$ is revealed after the recommending the item $\pi(i_t,t)$ to the new user. However, if this rating value is unknown in the testing data $Y$, we will replace it by zero. 

In practice, it is not always that the value $b^{*}_{t}$ is accessible, yet we can use the maximal value of the rating scale in recommendation systems, such as the number 5 in the Netflix data. Moreover, in some recommendation systems, we do not get the explicit values of $b_{\pi(i_t,t)}$. For this case, we must define another online evaluation measurement. The topic is out of the scope of this paper and we reserver it for the future study. 

\subsection{Results and Analysis}

\subsubsection{Results on dealing with missing values}
~\\
We filled the missing values in the base matrix $X$ by using three approaches described above. We then ran the A-LinUCB algorithm with $\alpha=0.001$, which is the best choice in our experiments, on the matries $Y$ of the Movielen, Netflix and Yahoo!Music data sets. Finally, we measured the cumulative regrets of the new user recommendation systems. The Table~\ref{table:missingValuePerformance} shows the obtained performances.

It can be seen that the Average approach works best with smallest cumulative regret. The Zero method is slightly worse than the Average on the Movielen data set only, yet it gained more efficiency in terms of running time. Supprisingly, the impute SVD and the alternative SVD provided worst resutls in our experiments.

Over all, we decided to use the Zero approach for the further comparisons because of its performances and efficency.
\begin{table*}
  \caption{Cumulative regrets for the new user recommendation system to compare three different approaches in dealing with missing values, namely by zero value, by average value, by impute SVD and by alternative SVD.}
  \centering
  \begin{tabular}{|l|c|c|c|c|}
    \hline
    Dataset & By zero & By average & By impute SVD & By alt.SVD\\
    \hline
    Movielen & 3008.19 & 3006.8 & 4872 & 4869.2\\
    \hline
    Netflix & 3680 & 3680 & 3856 & 3856.8\\
    \hline  
    Yahoo!Music & 989.56 & 989.56 & 1143.81 & 1512.35\\
    \hline
  \end{tabular}
  \label{table:missingValuePerformance}
\end{table*}

\subsubsection{Results on competing different methods}
~\\
As shown in Table~\ref{table:newUser} and Table~\ref{table:newItem}, obviously the random policy is much worse than the A-LinUCB in terms of the cumulative regrets. For example, in the cased of the new user recommendation system on the Yahoo!Music data set, the A-LinUCB algorithms provided a better performance than the random's result by 34\%.

\begin{table}
  \caption{Cumulative regrets for the \textbf{new user} recommendation system.}
  \centering
  \begin{tabular}{|l|c|c|c|}
    \hline
    Methods & Movielens & Netflix & Yahoo!Music \\
    \hline
    Random & 2024.26 & 3852 & 1514.30\\
    \hline
    Aver & 1255 & 3825 & 1125.15 \\
    \hline
    EGreedy & 1218.4 & 3758.66 & 1108.97 \\
    \hline  
    UCB & 1973.3 & 3850.86 & 1514.77 \\
    \hline
    EXP3 & 1868.4 & 3836.66 & 1276.42 \\
    \hline
     LinUCB & 1595.86 & 3732.86 & 1325.86 \\
    \hline
    A-LinUCB & 1069.8 & 3659.4 & 989.92 \\
    \hline
  \end{tabular}
  \label{table:newUser}
\end{table}

\begin{table}
  \caption{Running time of the \textbf{new user} recommendation system (in seconds).}
  \centering
  \begin{tabular}{|l|c|c|c|}
    \hline
    Methods & Movielens & Netflix & Yahoo!Music \\
    \hline
    Random & 93 &  15.16& 12.43\\
    \hline
    Aver & 117 & 21.9 & 18.44 \\
    \hline
    EGreedy & 113 & 22.12 & 18.57\\
    \hline  
    UCB & 48.96 & 22.39 & 18.77\\
    \hline
    EXP3 & 38.28 & 23.43 & 19.12\\
    \hline
     LinUCB & 20214 & 2361.6 & 2448 \\
    \hline
    A-LinUCB & 5184 & 654 & 612\\
    \hline
  \end{tabular}
  \label{table:newUserTime}
\end{table}

\begin{figure*}[ht!]
  \centering
  \begin{center}
    \subfigure[Movielens with size $6040\times 3952$. ]{\includegraphics[width=.3\textwidth]{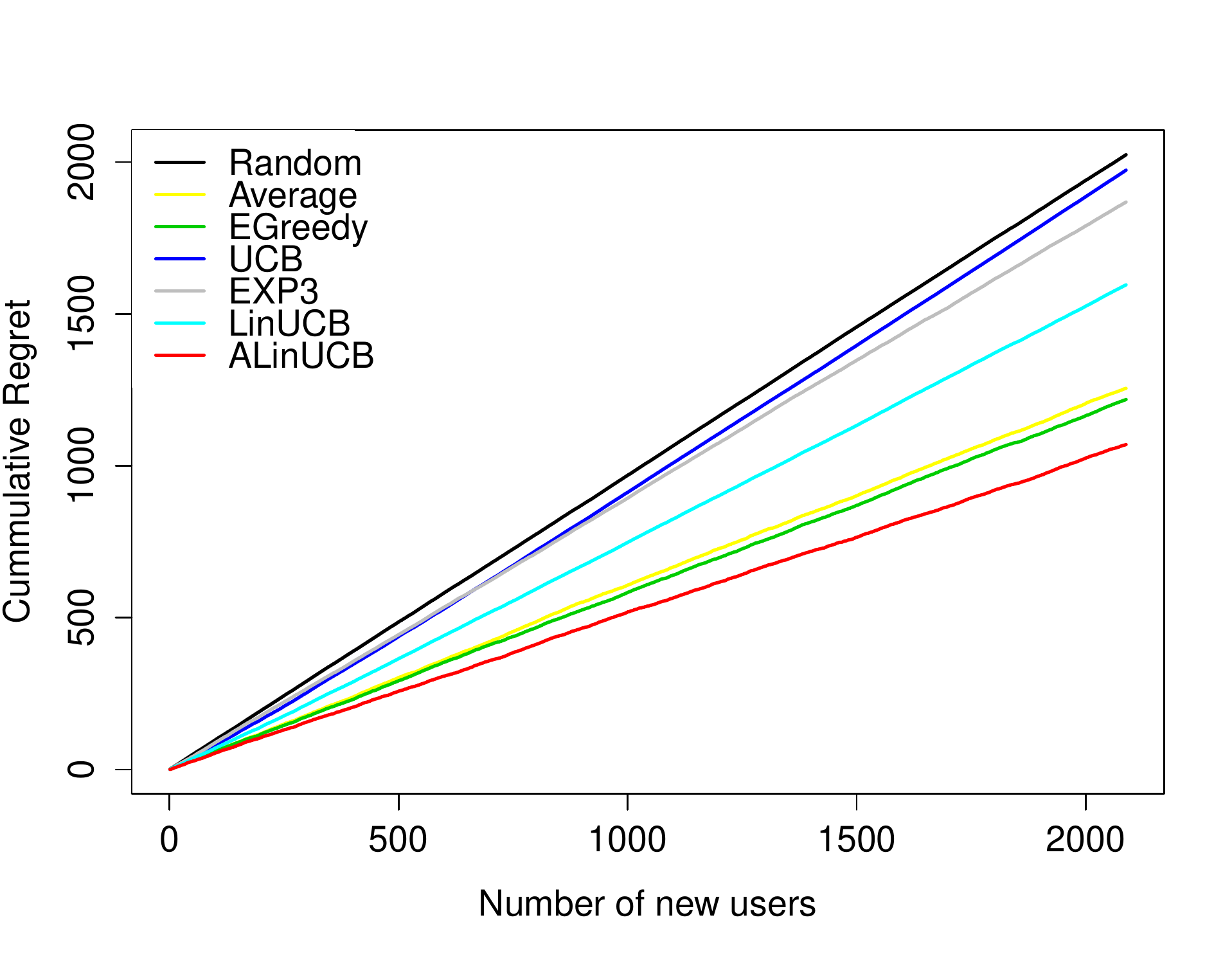}}
    \subfigure[Netflix with size $6423\times 1250$ ]{\includegraphics[width=.3\textwidth]{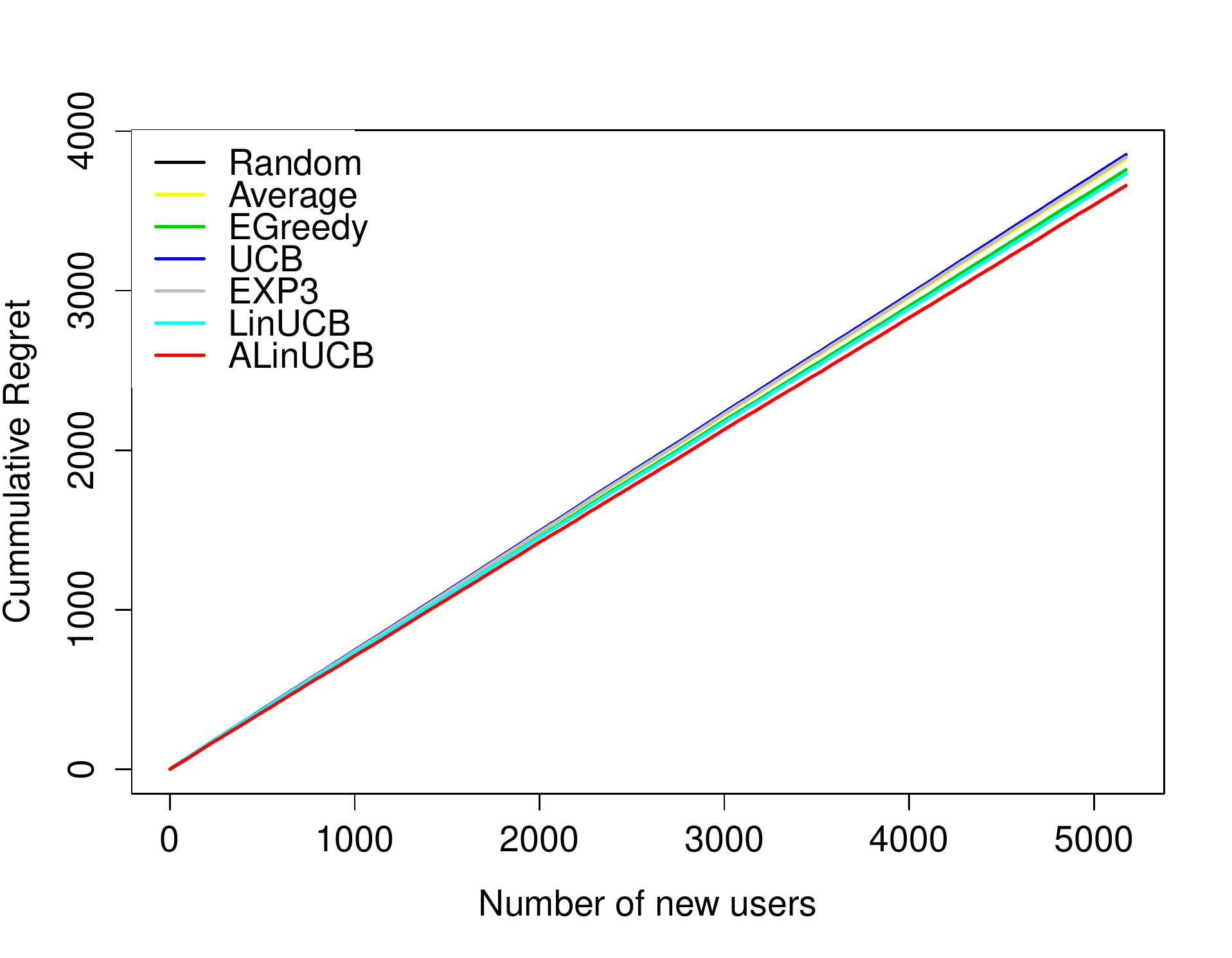}}
    \subfigure[Yahoo!Music with size $10,000\times 1000$]{\includegraphics[width=.3\textwidth]{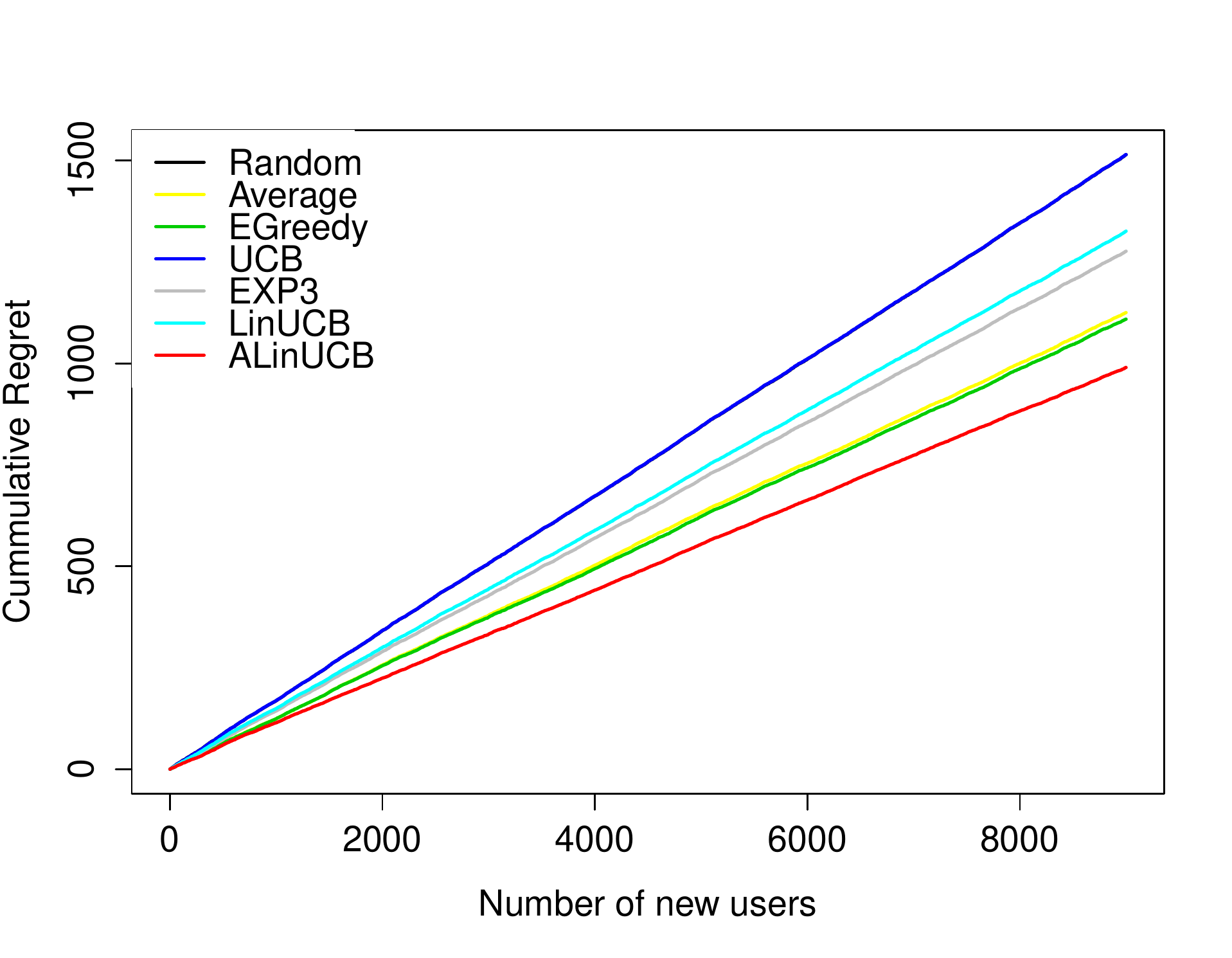}}
  \end{center}
  \caption{Cumulative regrets for the \textbf{new user} recommendation systems. Several curses are hidden on the graphs as the performances of these methods is almost the same.}
  \label{fig:newUser}
\end{figure*}

\begin{table}
  \caption{Cumulative regrets for the \textbf{new item} recommendation system.}
  \centering
  \begin{tabular}{|l|c|c|c|}
    \hline
    Methods & Movielens & Netflix & Yahoo!Music \\
    \hline
    Random & 2226.5 & 179.6 & 314.59\\
    \hline
    Aver & 1721.86 & 179.6 & 285 \\
    \hline
    EGreedy & 1664.46 & 178.06 & 273.81 \\
    \hline  
    UCB & 2221.46 & 179.93 & 313.75 \\
    \hline
    EXP3 & 2133.19 & 179.2 & 306.92 \\
    \hline
     LinUCB & 1715.73 & 173.46 & 273.24 \\
    \hline
    A-LinUCB & 1645.46 & 169.6 & 266.95 \\
    \hline
  \end{tabular}
  \label{table:newItem}
\end{table}

\begin{table}
  \caption{Running time of the \textbf{new item} recommendation system (in seconds).}
  \centering
  \begin{tabular}{|l|c|c|c|}
    \hline
    Methods & Movielens & Netflix & Yahoo!Music \\
    \hline
    Random & 35.36 & 0.94 & 12.26\\
    \hline
    Aver & 47.75 & 1.49 & 18.28 \\
    \hline	
    EGreedy & 45.29 & 1.46 & 18.41\\
    \hline  
    UCB & 35.62 & 1.43 & 18.67\\
    \hline
    EXP3 & 30.32 & 1.46 & 19.59\\
    \hline
     LinUCB & 2325.27 & 57.5 & 2866.68 \\
    \hline
    A-LinUCB & 547.75 & 13.8 & 360\\
    \hline
  \end{tabular}
  \label{table:newItemTime}
\end{table}
\begin{figure*}[ht!]
  \centering
  \begin{center}
    \subfigure[Movielens with size $1000\times 3952$]{\includegraphics[width=.3\textwidth]{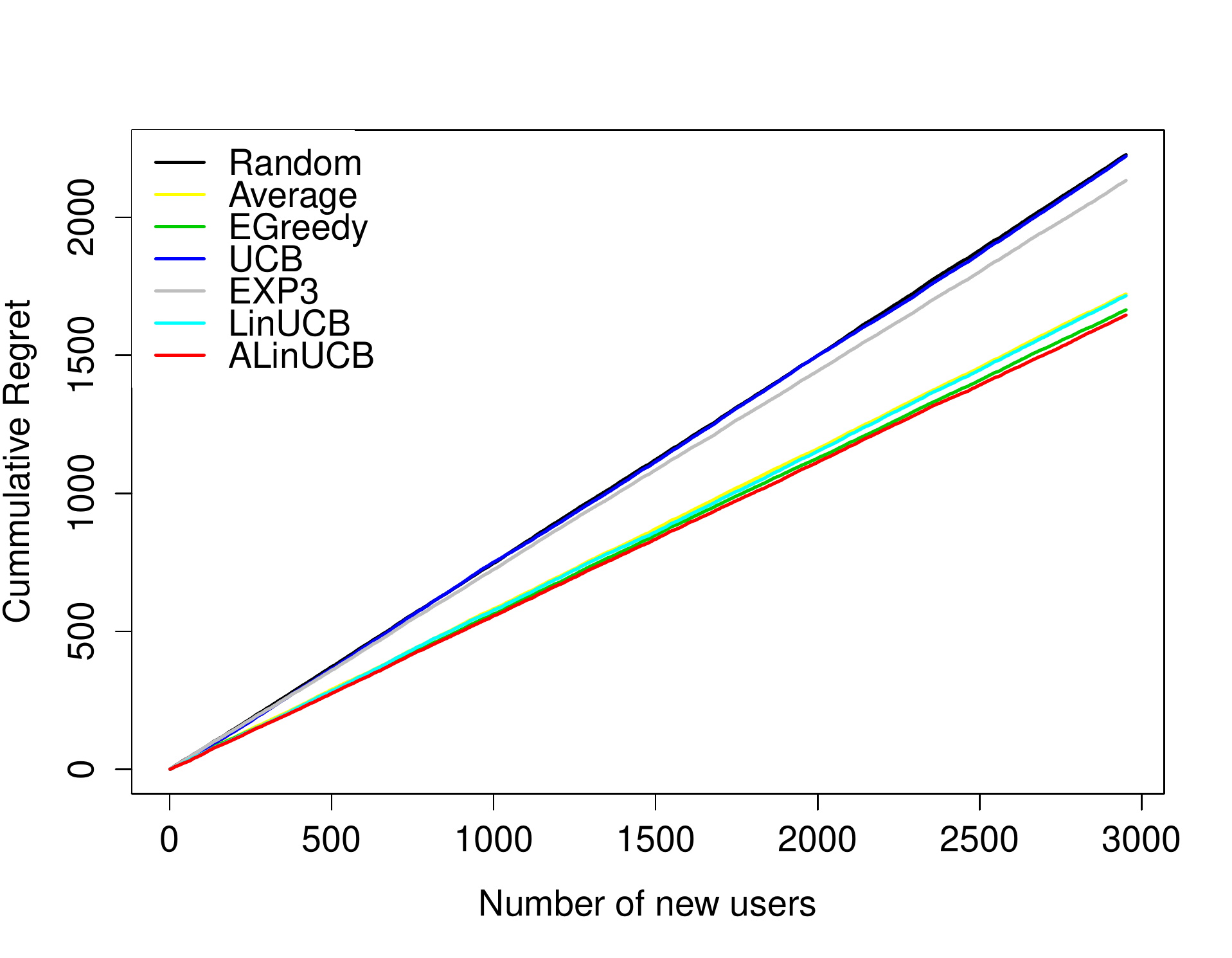}}
    \subfigure[Netflix with size $500\times 1250$]{\includegraphics[width=.3\textwidth]{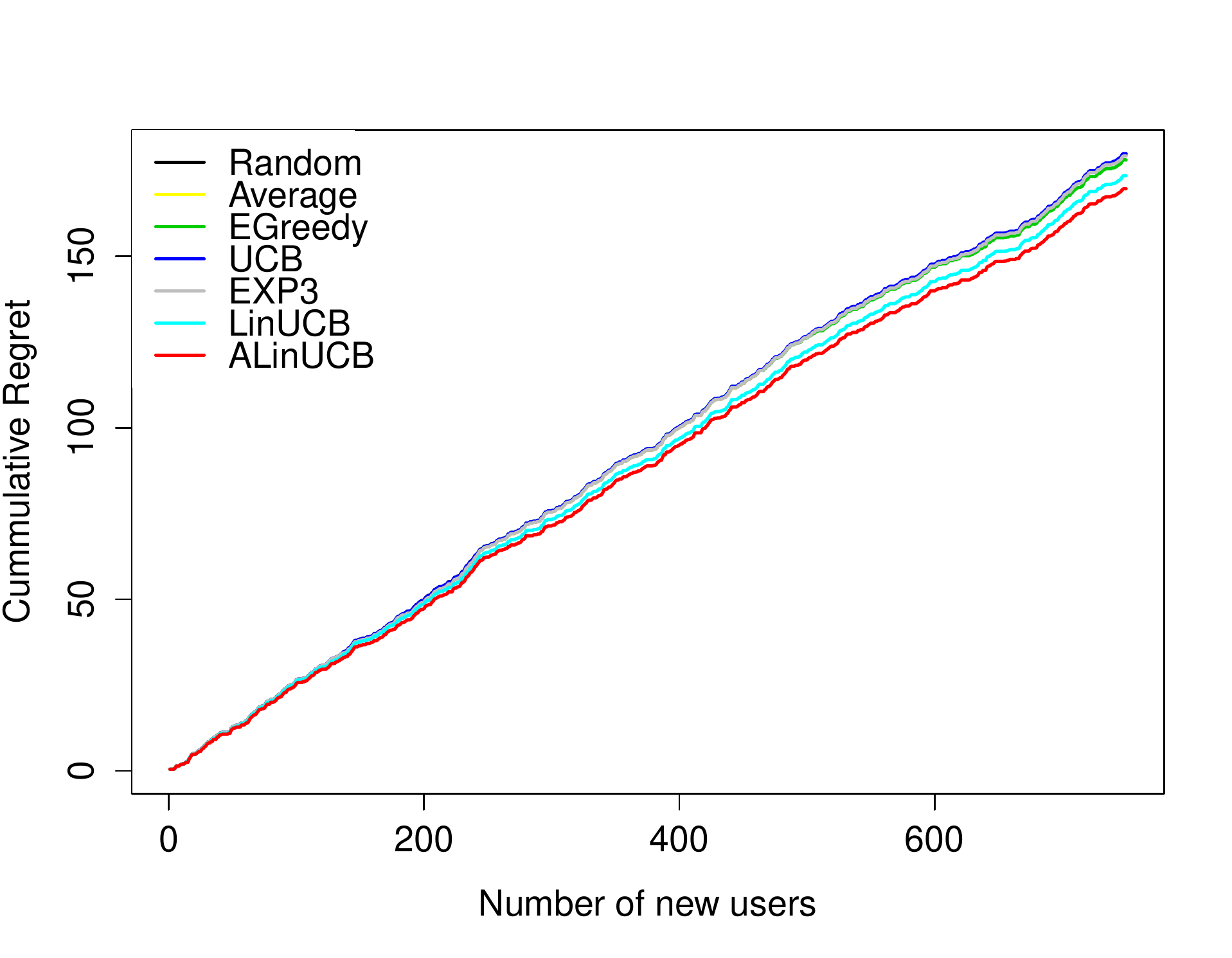}}
    \subfigure[Yahoo!Music with size $1000\times 10,000$]{\includegraphics[width=.3\textwidth]{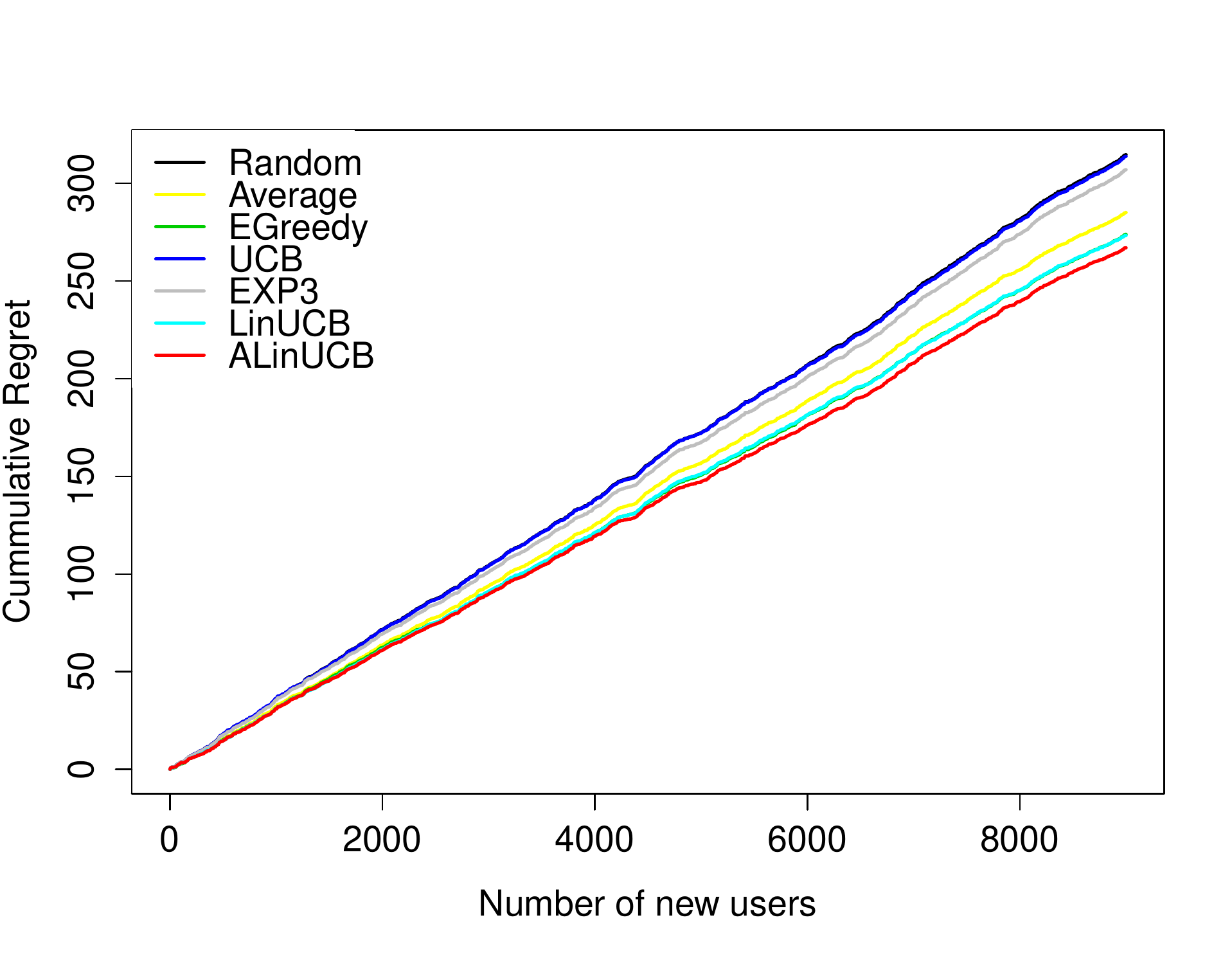}}
  \end{center}
  \caption{Cumulative regrets for the \textbf{new item} recommendation systems. Several curses are hidden on the graphs as the performances of these methods is almost the same.}
  \label{fig:newItem}
\end{figure*}
Clearly, with no use of the information about previous ratings the UCB and EXP3 algorithms
performed almost the same like the random strategy. Even the EXP3 is the worst choice in terms of
the cumulative regret and the running time. The EG algorithm is exceptional since its performance
is much better than the UCB and EXP3, yet still lower than the A-LinUCB. This can be explained by
the fact that an item, which is preferred by an enough number of users, will likely be a good
choice for the others. The greedy EG strategy has fully utilized this character of a recommendation
systems to provide good results. For instance, in the case of the new movie recommendation system
built on the Movielens data set, the EG policy performed very well with almost 10\% better than the
random strategy.

With the Thompson sampling (TS) policy, we got better results than the Random, UCB and EXP3. This illustrated that the information of previous ratings contributed to the better recommendation results. However, its performance is still lower than the EG and A-LinUCB. Therefore, we can conclude that the context information must be used in an appropriate way, which is the A-LinUCB as a sample in our experiments. 

It can be seen from the Table~\ref{table:A-LinUCB(0)vsA-LinUCB(0.001)} that the A-LinUCB algorithm with $\alpha=0$ performed much worse than the A-LinUCB when $\alpha=0.001$ on all three data sets. That means a little exploration helped to provide better results in solving cold-start recommendation systems. 

\begin{table}
  \caption{Cumulative regrets for new user recommendation system to
    compare between A-LinUCB ($\alpha=0$) with A-LinUCB
    ($\alpha=0.001$).}
  \centering
  \begin{tabular}{|l|c|c|c|}
    \hline
    A-LinUCB & Movielens & Netflix & Yahoo!Music \\
    \hline
    $\alpha=0$ & 3491.39 & 3857 & 1514.15 \\
    \hline
    $\alpha=0.001$ & 3008.19 & 3680 & 989.56 \\
    \hline
  \end{tabular}
  \label{table:A-LinUCB(0)vsA-LinUCB(0.001)}
\end{table}

%

\section{Related work}

The cold-start problem was readily identified as of the emergence of recommendation
systems~\cite{Schein:2002:MMC:564376.564421}. Along time, several solutions for these problems were
proposed. However, these approaches depend strongly on side information available on the users and
items, which is not always available. Therefore, it is very hard to build accurate recommendation
systems in practice, despite the strong activity on the subject. For example, the
work~\cite{Lashkari94collaborativeinterface} suggested an interview process with users to gather
more information about their preferences before the actual recommendations. Recently, a lot of
works have been conducted to improve the estimation speed of the parameters for new items or new
users by using hierarchy of items or various side
informations~\cite{DBLP:conf/nips/AgarwalCEMPRRZ08,contextualRecommendation,Agarwal:2009:SME:1526709.1526713}. The augmentation with information mined from social networks is also a common approach
today.

Another common strategy to mitigate the cold-start user problem is to gather demographic data. It is assume that users who share a common background also share a common taste in products. Examples include Lekakos and Giaglis \cite{lekakos07}, where lifestyle information is employed. This includes age, marital status and education, as well as preferences on eight television genres. Correlation between users are found by applying the Pearson correlation coefficient. The authors report that this approach is the most effective way of dealing with the cold-start user problem in sparse environments.

A similar thought underlies the work by Lam et al., \cite{lam08:_addres} where an \textit{aspect model} (see e.g. \cite{marlin04:_collab}) including age, gender and job is used. This information is used to calculate a probability model that classifies users into user groups and the probability how well liked an item is by this user group. 

Other examples of applying demographic information to mitigate the cold-start user problem exists, e.g. \cite{gao07:_person,agarwal09:_regres,park09:_pairw}. All the the solution above use similar demographic information; most commonly age, occupation and gender. Most of the solutions asked for less that five pieces of information. Even though five is a comparatively small number, the user must still answer these questions. Users do generally not like to answer a lot of questions, yet expect reasonable performance from the first interaction with the user \cite{zigoris06:_bayes}. 

Zigoris and Zhang \cite{zigoris06:_bayes}, suggests to use a two part Bayesian model, where the prior probability is based on the existing user population and \textit{data likelihood}, which is based on the data supplied by the user. Thus, when a new user enters the system, little is know about that user and the prior distribution is the main contributor. As the user interacts with the system the data data likelihood becomes more and more important. This approach performs well for cold-start users. Other similar approaches can by found in \cite{manavoglu03:_probab}, suggesting a Markov mixture model and \cite{xue09:_user}, who suggests a statical user language model that integrates an individual model, a group model and a global model.

In this paper, we consider this problem from a different
perspective. Though perfectly aware of the potential utility of side
information, we consider the problem without any side information,
only focussing on acquiring appetence of new users and appeal new
items as fast as possible with as few as possible ``bad'' recommendations.

\section{Conclusions and future study} 

The main focus of this paper is on cold-start problems in recommendation systems. We
have casted these problems as contextual-bandit problems and adapted LinUCB to solve
them. We have conducted performance analysis of the proposed A-LinUCB algorithms and compared it
with six different approaches: random policy, $\epsilon$-greedy, UCB, EXP3, Thompson sampling and
A-LinUCB($\alpha=0$). We have used three data sets from Movielens, Netflix and Yahoo!Music and the
performance of algorithms were measured by the cumulative regret. Our proposed A-LinUCB algorithms
have clearly demonstrated better results than all the others in both new user and new item
recommendation systems. As proposed in this paper, A-LinUCB requires no side information on users and items; A-LinUCB may be extended to take advantage of such sources of information; this is left as future work. 

There are two main directions for future works: first, it would be interesting to extend the proposed framework to solve the cold-start system problem, where we have completely new users and new items. Second, we plan to study another forms of the cumulative regret for the case when only implicit feedbacks of users are available, such as clicks or browsing. 
\bibliographystyle{plain}
\bibliography{ref}

\end{document}